\documentclass[twocolumn,showpacs,preprintnumbers,amsmath,amssymb,aps]{revtex4}

\usepackage{graphicx}
\usepackage{dcolumn}
\usepackage{bm}
\usepackage[colorlinks,linkcolor=blue,anchorcolor=blue,citecolor=blue]{hyperref}
\usepackage[caption=false,font=normalsize,labelfont=sf,textfont=sf]{subfig}
\usepackage{enumerate}
\usepackage{enumitem}
\usepackage{multirow}
\usepackage{booktabs}
\usepackage{amsmath,amsfonts}
\usepackage{bm}
\usepackage[ruled]{algorithm2e}
\usepackage{amsthm}

\newtheorem{lemma}{Lemma}

\begin{document}


\title{Quantum Convolutional Neural Network with Flexible Stride}

\author{Kai  Yu$^{1,2}$}
\author{Song Lin$^1$}%
\thanks{Corresponding author. Email address: lins95@fjnu.edu.cn}
\author{Bin-Bin Cai$^{1,2}$}
\thanks{Corresponding author. Email address: cbb@fjnu.edu.cn }
\affiliation{%
{$^1$College of Computer and Cyber Security, Fujian Normal University, Fuzhou 350117, China}}
\affiliation{{$^2$Digital Fujian Internet-of-Things Laboratory of Environmental Monitoring, Fujian Normal University, Fuzhou 350117, China}
}%


\date{\today}

\begin{abstract}
Convolutional neural network is a crucial tool for machine learning, especially in the field of computer vision. Its unique structure and characteristics provide significant advantages in feature extraction. However, with the exponential growth of data scale, classical computing architectures face serious challenges in terms of time efficiency and memory requirements. In this paper, we propose a novel quantum convolutional neural network algorithm. It can flexibly adjust the stride to accommodate different tasks while ensuring that the required qubits do not increase proportionally with the size of the sliding window. First, a data loading method based on quantum superposition is presented, which is able to exponentially reduce space requirements. Subsequently, quantum subroutines for convolutional layers, pooling layers, and fully connected layers are designed, fully replicating the core functions of classical convolutional neural networks. Among them, the quantum arithmetic technique is introduced to recover the data position information of the corresponding receptive field through the position information of the feature, which makes the selection of step size more flexible. Moreover, parallel quantum amplitude estimation and swap test techniques are employed, enabling parallel feature extraction. Analysis shows that the method can achieve exponential acceleration of data scale in less memory compared with its classical counterpart. Finally, the proposed method is numerically simulated on the Qiskit framework using handwritten digital images in the MNIST dataset. The experimental results provide evidence for the effectiveness of the model.
\end{abstract}

\maketitle


\section{\label{sec:1}Introduction}
\par Quantum mechanical properties, such as quantum superposition and quantum entanglement, are utilized in quantum information processing (QIP) \cite{nielsen2010quantum} to overcome the limitations of classical computing. And it has the potential to significantly increase computational speed for certain problems \cite{grover1997,Shor1999}. QIP is also applied to machine learning tasks \cite{biamonte2017,dunjko2018}, and some promising breakthroughs have been achieved in recent years. On one hand, purely quantum algorithms that rely entirely on quantum gates for evolution, such as quantum linear equation solving \cite{HHL,wossnig2018}, quantum dimensionality reduction \cite{congLDA2016,YCH2019,YK2023,feng2024quantum}, and quantum regression \cite{YCH2021IEEE, chen2022}. These algorithms leverage quantum universal gates to replicate the evolution of corresponding classical processes and achieve speedup. Another significant development in quantum machine learning is the variational quantum algorithm (VQA), which combines low-depth parameterized quantum circuits with classical optimization algorithms to accomplish computational tasks. It is more suitable for current noisy intermediate-scale quantum (NISQ) devices \cite{bharti2022noisy, huang2023near} and has been successful in a variety of tasks \cite{Li2023VQA,Wang2023VQA,song2024quantum,Miao2024VQA}.
\par Convolutional neural network (CNN) is a crucial branch of machine learning, known for its ability to extract data features hierarchically \cite{lecun1998,arbib2003}. It has many interesting applications in computer vision. However, in the era of big data, a CNN consumes significant computational resources to handle vast amounts of data. To overcome this challenge, a solution is to apply QIP techniques to CNN. In 2019, Cong et al. proposed a VQA for feature extraction of CNN using quantum parameterized circuits \cite{cong2019QCNN}. It not only has a logarithmic reduction in the input space, but also can be effectively trained and implemented on NISQ devices. Subsequently, scholars proposed a series of quantum variational CNN algorithms \cite{henderson2020, liu2021, hur2022}. These algorithms design a variety of optimization objectives and introduce the corresponding variational circuits, which effectively implement the core functions of CNN while demonstrating the potential advantages of QIP. However, the interpretability of VQA algorithm on Hilbert space of exponential dimension deserves further exploration \cite{mcclean2018barren}.

\par To fully exploit the potential that QIP may offer, researchers also explored pure quantum algorithms for the computation of functions related to CNN. Two relevant works are introduced, which enable CNN to be implemented based on frequently used quantum gate operations. Specifically, Kerenidis et al. proposed a quantum algorithm for deep CNN \cite{kerenidis2019QCNN}. This algorithm expands the region passed by each filter into a vector, and explores a quantum algorithm for calculating the inner product of two vectors to complete feature extraction. And Li et al. proposed a quantum CNN with a specific architecture, where both the stride and the sliding window size are equal and powers of two \cite{li2020QCNN}. It fully leverages the parallel advantages of the quantum paradigm in both storage and computation, achieving exponential acceleration compared with the classical counterpart. These algorithms have undergone rigorous theoretical validation, which showcases substantial quantum advantages on ideal quantum computers. However, ensuring that the quantum CNN algorithm can flexibly adjust steps without the qubit count increasing with the sliding window size remains a challenging problem. Therefore, the quantum CNN algorithm for general scenarios is worth further research.

\par In this paper, a quantum convolutional neural network algorithm with flexible stride (QCNNFS) is proposed. We first fully utilize quantum coherence, which allows the QCNNFS model to flexibly choose the strides to balance the needs of feature extraction and receptive field. The original data is loaded into the quantum system using an analog coding method, which is easy to implement. Subsequently, a series of quantum circuits are constructed to realize the corresponding quantum evolution based on the frequently used quantum gates. We also exploit the swap test \cite{buhrman2001} and quantum amplitude estimation (QAE) \cite{brassard2002} techniques to compute the inner product between the filter parameters and the receptive field data, thereby extracting the data features of the current receptive field in parallel. Furthermore, a unitary operator is designed for information interaction, when convolution layer and pooling layer are connected. The inverse process of quantum evolution is used to disentanglement, increasing the degree of freedom of the system and improving the reutilization of qubits in the QCNNFS model. Compared with the classical counterpart, the proposed quantum process has exponential acceleration in the data scale as revealed by the time complexity analysis. The memory complexity analysis demonstrates that the proposed method requires fewer qubits than other quantum CNN algorithms with quantum speedup. At the same time, the space needed by QCNNFS does not significantly increase with the size of the sliding window. Finally, numerical simulations utilizing handwritten digit images from the MNIST dataset validate the effectiveness of the model and underscore the critical role of stride flexibility.
\par The remainder of this paper is organized as follows. The classical CNN and four quantum subroutines are reviewed in Sec. \ref{sec:2}. In Sec. \ref{sec:3}, the implementation scheme for QCNNFS is presented. In Sec. \ref{sec:4}, the time complexity and the memory complexity of QCNNFS are discussed. Then, the experiments about the proposed method are performed in Sec. \ref{sec:5}. In Sec. \ref{sec:6}, we give the conclusion of our work.
\section{\label{sec:2}Preliminaries} 

\par In this section, we will briefly review the classical CNN in Sec. \ref{sec:2.1}. And some interesting quantum subroutines are introduce in Sec. \ref{sec:2.2}.

\subsection{\label{sec:2.1} Review of the classical CNN}

\par
\begin{figure}
        \centering
        \subfloat[]{
            \includegraphics[width=0.48\textwidth]{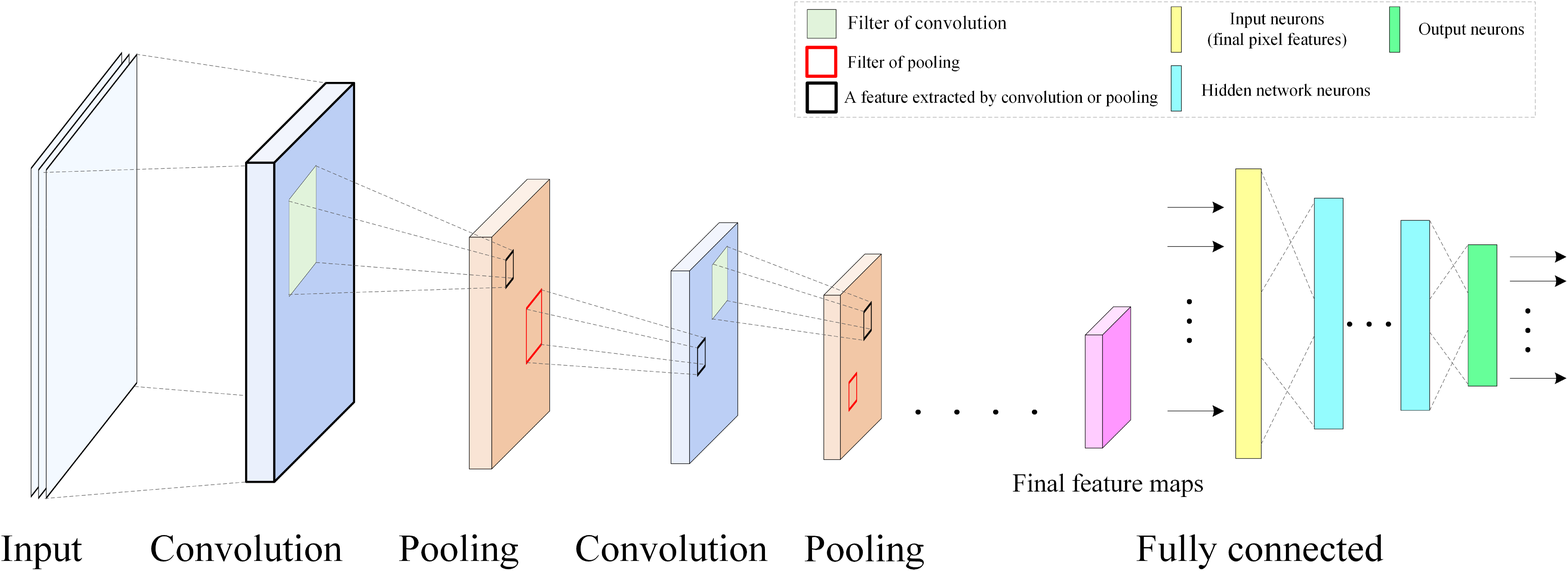}
        \label{fig:1a}
        }
        \newline
        \subfloat[]{
            \includegraphics[width=0.48\textwidth]{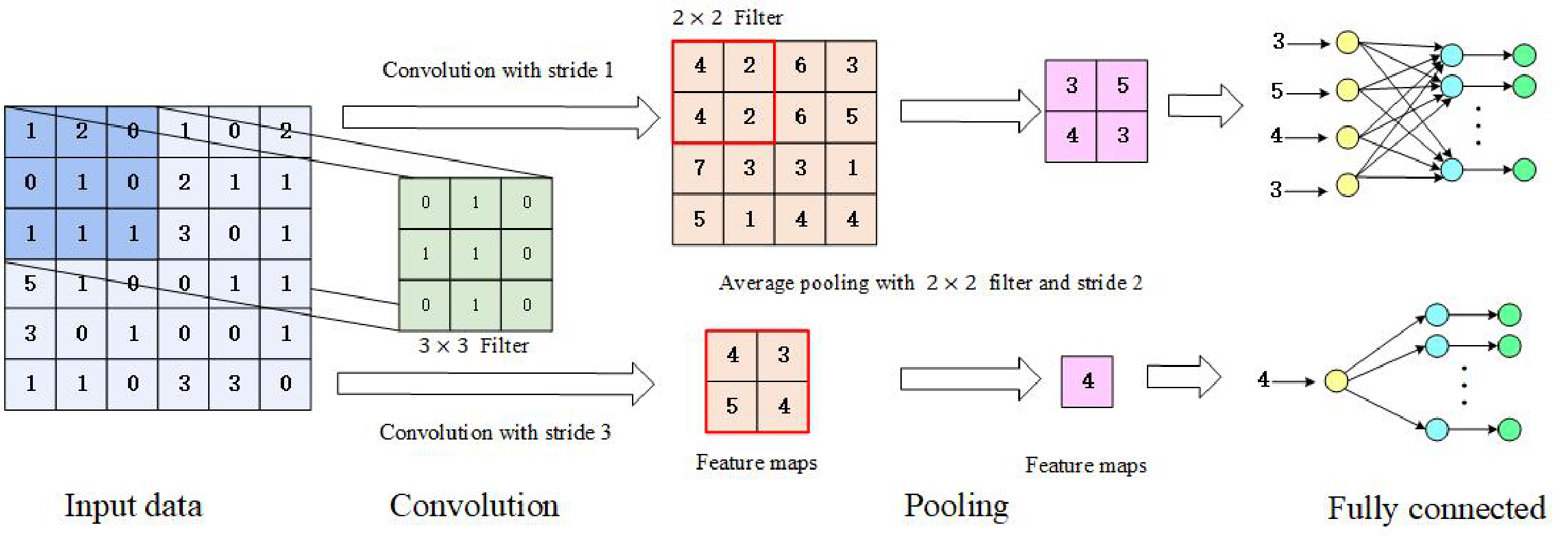}
        \label{fig:1b}
        }
        \caption{(a) A visualization of CNN framework. (b) An example of CNN operation with different stride settings.}
        \label{fig:1}
    \end{figure}

 A CNN generally consists of three layers, that is, convolution layers, pooling layers, and fully connected layers at the end. And the convolution layers and pooling layers may be repeated in any order. A visualization of the basic CNN framework is shown in Fig. \ref{fig:1a}. The functions of each layer are as follows.

\par \textbf{1) Convolution layer.} The operation of the convolutional layer is the core of CNN, primarily used for feature extraction. Its operation is that the convolution kernel (filter) slides over the input data with a predefined stride $s$ $(s>0,s\in \mathbb{Z})$ until it captures all data information. With each slide, a filter covers a local region of the input data, known as the receptive field. In each receptive field, an inner product of the data and the filter weights are calculated to generate a feature. These features collectively form feature maps, capturing the characteristics of the input data across different local areas. Specifically, the extracted pixel feature $r_{{x^{\prime}},{y^{\prime}}}^{\prime}$ at position $(x^{\prime},y^{\prime})$ in the feature map can be calculated by $r_{{x^{\prime}},{y^{\prime}}}^{\prime} = \sum_{a,b=0}^{N-1} {k_{a,b}} {r_{x,y}}$, where the weights ${k_{a,b}}$ from a convolution kernel $K \in \mathbb{R}^{N \times N}$, and ${r_{x ,y}}$ ($x=x^{\prime}s+a, y=y^{\prime}s+b$) is the data in the current receptive field.

\par \textbf{2) Pooling layer.} The functions of the pooling layer are mainly reduction and feature screening. Similarly, the pooling operation employs a sliding window to slide over the feature map with a preset stride $s^{\prime}$ $(s^{\prime}>0,s^{\prime}\in \mathbb{Z})$, covering a local region of the feature map with each slide. A popular polling operation is average pooling, which takes the average value over the region covered by the filter. Let a filter be represented as a $N^{\prime} \times N^{\prime}$ matrix, then the data sampling about its coverage region $\mathcal{R}$ is $r^{\prime\prime} = \frac{1}{{N^{\prime}}^{2}}\sum_{x^{\prime},y^{\prime}\in {\mathcal{R}}} {r_{x^{\prime},y^{\prime}}^{\prime}}$. These sampled values form a new feature map that preserves the salient features of the input data and reduces unnecessary information. In this way, not only the computational complexity and memory requirements are effectively reduced, but also the risk of overfitting is reduced.

\par \textbf{3) Fully connected layers.} The fully connected layer is the last module of a CNN, which connects pixel features to output nodes with weights. It is used to learn the features of the input data and to produce corresponding predictions. In classification tasks, each node represents a category label whose value is obtained from a linear combination of input elements and weights. The values of these nodes can be used to describe the confidence level that the input data belongs to the corresponding category.

\par An illustrative example of a simple CNN operation is demonstrated in Fig. \ref{fig:1b}. Furthermore, we give the time complexity required for each layer of a classical CNN. In the convolution layer, the time complexity of building a feature map is $O(M^2N^2)$, where $M^2$ is the size of the input data and $N^2$ is the number of weights in the convolutional kernel. Similarly, the operations of the pooling layer are executed in time $O({M^{\prime}}^2{N^{\prime}}^2)$. Here, ${M^{\prime}}^2$ is labeled as the number of data in the input feature map and ${N^{\prime}}^2$ is denoted as the size of the pooling filter. The time complexity of the fully connected layer depends on the number $\bar{M}$ of input features and the number $K$ of output nodes. It requires a time cost of $O(\bar{M}K)$.

\subsection{\label{sec:2.2} Quantum subroutines}

\par To facilitate the understanding, four quantum subroutines are introduced, which will be used to construct the specific procedure of QCNNFS.

\begin{lemma}[Quantum random access memory algorithm \cite{QRAM, Giovannetti2008}]
Let an $M$-dimensional vector $\bm{\vartheta} = ( {\vartheta_{0}, \vartheta_{1}, \cdots, \vartheta_{M - 1}} )$ be the provided data, which is already stored in the quantum random access memory (QRAM). There is an oracle $\mathcal{O}_{QRAM}$ to implement quantum base encoding of the data, $\mathcal{O}_{QRAM}\sum_{m = 0}^{M - 1} |m \rangle |0 \rangle  = {\sum_{m = 0}^{M - 1} |m\rangle} |\vartheta_{m}\rangle,$ in time $O[\mathrm{poly}\log M]$.
\label{lem:QRAM}
\end{lemma}
\par To obtain the probability estimate of the inner product required for the quantum CNN algorithm, the quantum amplitude estimation subroutine developed in Ref. \cite{brassard2002} will be utilized. This subroutine is described in the following lemma.
\begin{lemma}[Quantum amplitude estimation \cite{brassard2002}]
Suppose that a unitary operator $U$ can implement the transformation $U|0\rangle^{\otimes{L}} \rightarrow \sin{\theta} |\psi_{0},0\rangle + \cos{\theta} |\psi_{1},1\rangle$ in time $O(T)$, where $|\psi_{0}\rangle$ is some desired state and $L$ is the number of qubits. Then, the unitary operator $Q = -US_{1}U^{\dagger}S_{0}$ can be constructed, where $S_{1} = I-2(|0\rangle \langle0|)^{\otimes{L}}$, $S_{0} = I^{\otimes{L-1}} \otimes (I-2|0\rangle \langle 0|)$. Moreover, there is a quantum algorithm to estimate the amplitude $\theta$ to be $\tilde{\theta}$ within the error $\epsilon$, in which the time complexity is $O(T/{\epsilon})$.
\label{lem:QAE}
\end{lemma}
\par As stated in Lemma \ref{lem:QAE}, the implementation of quantum amplitude estimation relies on a unitary operation $U$, which generates a specific quantum entangled state. In certain cases, the inverse operation $U^{\dagger}$ of the unitary operator $U$ can be performed to disentangle, such as $U^{\dagger} (\sin{\theta} |\psi_{0},0\rangle + \cos{\theta} |\psi_{1},1\rangle ) \rightarrow |0\rangle^{\otimes{L}}$. The following lemma presents a unitary process, whose generated quantum state's probability distribution at a certain moment can be used to estimate the required inner product.
\begin{lemma}[Swap test \cite{buhrman2001}]
Given two vectors $r$, $w \in \mathbb{R}^{N}$. And their corresponding quantum states $|r\rangle$, $|w\rangle$ can be prepared by the unitary operators $U_r$ and $U_w$, respectively. There exists a procedure to produce an entangled state $(|r\rangle + |w\rangle)|0\rangle + (|r\rangle - |w\rangle) |1\rangle$. The inner product $\left\langle r \middle| w \right\rangle$ can then be estimated based on the probability of this quantum entangled state being in state $|0\rangle$ or $|1\rangle$.
\label{lem:ST}
\end{lemma}
\par We also require some arithmetic operations to implement various function computations within a quantum system. An efficient arithmetic computation tool for the basis-encoded states is provided by quantum arithmetic operations, as stated in the following lemma.
\begin{lemma}[Quantum arithmetic operations \cite{ruiz2017, zhouss2017}]
Let $\alpha$, $\beta$ be two basis-encoded quantum states with $L$ qubits. The arithmetic operation $|\alpha\rangle |\beta \rangle \rightarrow |\alpha \rangle | \alpha + \beta \rangle$ or $|\alpha \rangle |\beta \rangle \rightarrow |\alpha \rangle |\alpha \beta \rangle$ can be calculated in time $O(\mathrm{poly}(L))$, by a quantum circuit.
\label{lem:QAC}
\end{lemma}
%
%
\section{\label{sec:3}QCNNFS model}
\par In this section, the implementation scheme for the QCNNFS model will be presented. First, Sec. \ref{sec:3.0} gives the method to prepare the classical information in a quantum system. The quantum circuit to implement the convolution layer is presented in Sec. \ref{sec:3.1}. Then, the way to realize the operations of the pooling layer in a quantum system is proposed in Sec. \ref{sec:3.2}. Finally, the quantum fully connected layer is designed in Sec. \ref{sec:3.3}.
\subsection{\label{sec:3.0}Quantum preparation of classical information}
\par Inspired by classical image representations, pixel information can be integrated into a matrix
     \begin{equation}
        G = \begin{bmatrix}
        r_{0,0} & \cdots & r_{0,M - 1} \\
        \vdots & \ddots & \vdots \\
        r_{M - 1,0} & \cdots & r_{M - 1,M - 1}
        \end{bmatrix}
     \label{eq:2}
     \end{equation}
for a gray image, where $r_{x,y} \in \lbrack 0,255\rbrack$ represents the pixel value at position $(x,y)$ with $x = 0, 1, \cdots, M-1$ and $y = 0, 1, \cdots, M-1$. Without loss of generality, the pixel values are normalized to
     \begin{equation}
     {\tilde{r}}_{x,y} = \frac{r_{x,y}}{\left\| G \right\|_{F}}.
     \label{eq:3}
     \end{equation}
Here, ${\left\| G \right\|_{F}}$ is labeled as the Frobenius norm of the matrix $G$.
\par The prerequisite for effectively applying quantum algorithms is encoding classical information into quantum systems. A rotation operator \cite{nielsen2010quantum} about the $y$-axis is performed to encode classical information into the amplitude of the computational basis state. The quantum state of the pixel value at position $(x,y)$ can be described as
     \begin{equation}
     R ( \vartheta_{x,y} ) | 0\rangle = | {\tilde{r}}_{x,y} \rangle,
     \label{eq:4}
     \end{equation}
where $R\left( \vartheta_{x,y} \right) = \begin{bmatrix}
{\cos\vartheta_{x,y}} & {- {\sin\vartheta_{x,y}}} \\
{\sin\vartheta_{x,y}} & {\cos\vartheta_{x,y}}
\end{bmatrix}$, $| {\tilde{r}}_{x,y} \rangle = {\cos\vartheta_{x,y}} |  0 \rangle  + {\sin\vartheta_{x,y}} |  1 \rangle $, and $\vartheta_{x,y} = \arccos{\tilde{r}}_{x,y}$. Generally, for an image with $M^{2}$ pixels, using this method to encode the image requires a time cost of $O({M^{2}})$.
\par
\begin{figure}
        \centering
        \includegraphics[width=0.48\textwidth]{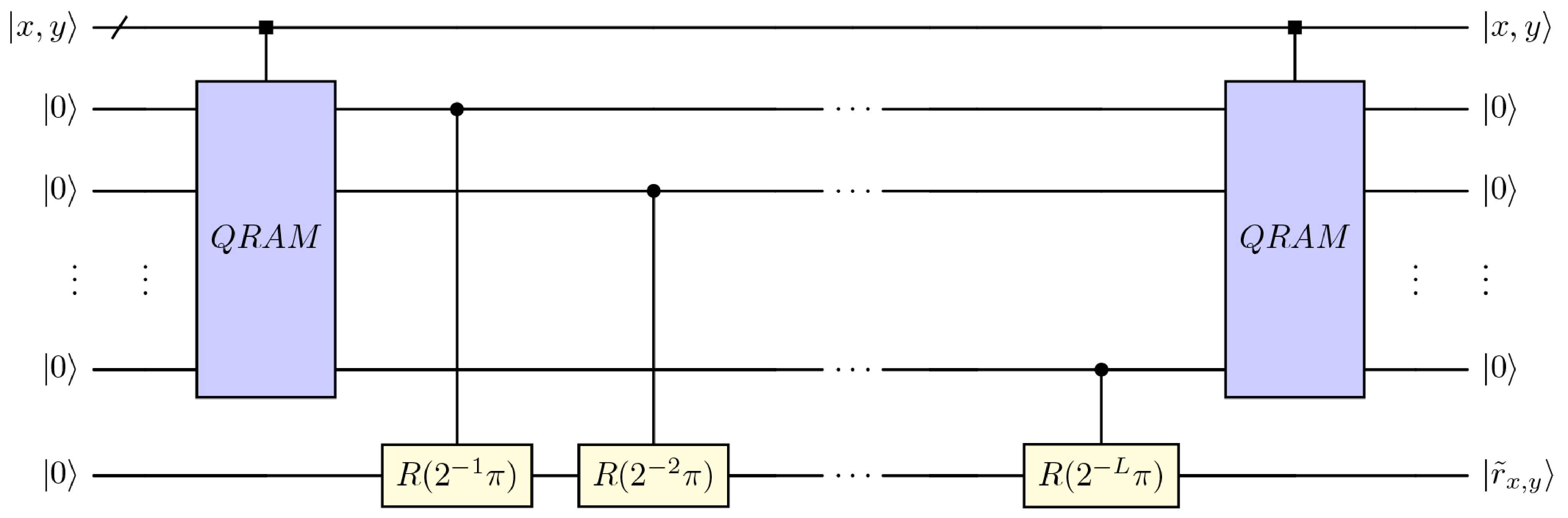}
        \caption{The circuit for the operation $\tilde{\mathcal{O}}$.}
    \label{fig:O}
    \end{figure}

 However, if the angles $\vartheta_{x,y}$ ($x = 0, 1, \cdots, M-1$, $y = 0, 1, \cdots, M-1$) related to pixels are stored in the QRAM, we can load the image information into the quantum system in polynomial logarithm time, according to Lemma \ref{lem:QRAM}. Specifically, $\vartheta_{x,y}$ can be preprocessed as $\tilde{\vartheta}_{x,y} = \vartheta_{x,y} / \pi$. And its binary representation within the error $2^{-L}$ is ${\tilde{\vartheta}}_{x,y} = {\tilde{\vartheta}}_{x,y}^{1} \cdots {\tilde{\vartheta}}_{x,y}^{L}$, i.e., $\vartheta_{x,y} \approx {\sum_{l = 1}^{L}{\tilde{\vartheta}}_{x,y}^{l}}2^{- l}\pi$. With the help of $\mathcal{O}_{QRAM}$, the information can be encoded into the quantum state ${\sum_{x,y = 0}^{M - 1} | {x,y} \rangle } | {\tilde{\vartheta}}_{x,y} \rangle$ in time $O[\mathrm{poly}\log M^2]$. Then, the implementation of operator $R(2^{-l}\pi)$ is controlled by the $l$-th qubit of $| {\tilde{\vartheta}}_{x,y} \rangle$, we can approximate the operation $R ( \vartheta_{x,y} ) \approx {\prod_{l = 1}^{L}{R ( {{\tilde{\vartheta}}_{x,y}^{l}2^{- l}\pi} )}}$. In this case, the pixel information can be encoded as a quantum state ${\sum_{x,y = 0}^{M - 1} | {x,y} \rangle} | {\tilde{r}}_{x,y} \rangle$ in time $O[\mathrm{poly}\log M^2+ L]$. This method is also applicable for encoding individual elements within the data, with the corresponding operator defined as
     \begin{equation}
     \tilde{\mathcal{O}} | {x,y} \rangle |  0 \rangle = | {x,y} \rangle |  {\tilde{r}}_{x,y} \rangle.
     \label{eq:5}
     \end{equation}
The quantum circuit implementing $\tilde{\mathcal{O}}$ is given in Fig. \ref{fig:O}. It is worth noting that this operator is typically utilized during the subsequent information loading process.
\subsection{\label{sec:3.1}Quantum convolution layer}
\par Without loss of generality, considering an input image pixel information matrix $G \in \mathbb{R}^{M \times M}$, a convolution kernel $W \in \mathbb{R}^{N \times N}$ (with elements $w_{i,j}$ at position $(i,j)$), and a stride $s$ $(s>0,s\in \mathbb{Z})$. As described in Sec. \ref{sec:3.0}, we assume that their relevant information is already stored in the QRAM \cite{QRAM}. The operation of the quantum convolution layer and the quantum state transformation are given in the following.
\par (S1.0) Initializing a quantum system in the state
     \begin{equation}
     \frac{1}{M^{\prime}N} {\sum\limits_{x^{\prime}, y^{\prime} = 0}^{M^{\prime} - 1}|  {x^{\prime},y^{\prime}} \rangle_{c_1}}{\sum\limits_{i,j = 0}^{N - 1} | {i,j} \rangle_{c_2}}
     \label{eq:6}
     \end{equation}
to generate positional space for data information. The registers marked with different subscripts serve distinct purposes. For example, quantum register ${c_1}$ is used to store position information in the feature graph generated by the convolution, while register ${c_2}$ holds the position index of the convolution kernel. $M^{\prime}$ represents the dimension of the feature matrix generated by the convolution operation, which can be calculated as $M^{\prime} = \frac{M - N}{s} + 1$. Additionally, we denote $|{x^{\prime},y^{\prime}}\rangle = |x^{\prime} \rangle | y^{\prime} \rangle$ and $|{i,j}\rangle = |i \rangle | j \rangle$.

\par (S1.1) A new register $c_3$ in state $| 0 \rangle^{\bigotimes {\lceil {\log{s}} \rceil} }$ is introduced. Then, the $X^{s^{l}}$ $(l =1, \cdots, {\lceil \log{s} \rceil})$ operations are performed on it based on the binary representation $s^{1} s^{2} \cdots s^{\lceil \log{s} \rceil}$ of the stride $s$. The quantum system sate becomes into
     \begin{equation}
     \frac{1}{M^{\prime}N}{\sum\limits_{x^{\prime},y^{\prime} = 0}^{M^{\prime} - 1} |  {x^{\prime},y^{\prime}} \rangle _{c_{1}}}{\sum\limits_{i,j = 0}^{N - 1} | {i,j} \rangle _{c_{2}}} |  s \rangle _{c_{3}}.
     \label{eq:7}
     \end{equation}

\par (S1.2) A state $| 0 \rangle^{\bigotimes 2{\lceil{\log{M}}\rceil}}$ is initialized in the register $c_4$. Based on the fact that the relationship $x=x^{\prime}s+i$, $y=y^{\prime}s+j$ between the feature position information $(x^{\prime}, y^{\prime})$ generated by the convolution and the position information $(x,y)$ of the input image, the quantum arithmetic operations described in Lemma \ref{lem:QAC} can be applied to obtain the state
     \begin{equation}
     \frac{1}{M^{\prime}N}{\sum\limits_{x^{\prime},y^{\prime} = 0}^{M^{\prime} - 1} |  {x^{\prime},y^{\prime}} \rangle _{c_{1}}}{\sum\limits_{i,j = 0}^{N - 1} | {i,j} \rangle _{c_{2}}} |  s \rangle _{c_{3}} |x,y \rangle_{c_4}.
     \label{eq:8}
     \end{equation}

\par (S1.3) The inverse operations of step (S1.1) are performed on register $c_3$. For simplicity, we denote $| m^{\prime} \rangle = |  {x^{\prime},y^{\prime}} \rangle$, $ |n\rangle = | {i,j} \rangle$, and $ | m \rangle = |x,y \rangle$. A simple example is that $|00,01\rangle$ can be relabeled as $|0001\rangle$. Thus, the state of quantum system can be expressed as
     \begin{equation}
     \frac{1}{{M^{\prime}}N}{\sum\limits_{{m^{\prime}} = 0}^{{M^{\prime}}^{2} - 1}} |  m^{\prime} \rangle _{c_{1}} \sum\limits_{n = 0}^{N^{2} - 1} | {n} \rangle _{c_{2}} |m \rangle_{c_4}.
     \label{eq:9}
     \end{equation}
\par (S1.4) Appending three registers in the state $|0\rangle_{c_5} |0\rangle_{c_6} |0\rangle_{c_7}$. A $Hadamard$ gate is applied on the register $c_{5}$ to obtain
     \begin{equation}
     \frac{1}{\sqrt{2}{M^{\prime}}N}{\sum\limits_{{m^{\prime}} = 0}^{{M^{\prime}}^{2} - 1}} |  m^{\prime} \rangle _{c_{1}} \sum\limits_{n = 0}^{N^{2} - 1} | {n} \rangle _{c_{2}} |m \rangle_{c_4} ( |0\rangle + |1\rangle )_{c_{5}} |00\rangle_{c_{6,7}}.
     \label{eq:10}
     \end{equation}
Here, $| \cdot \rangle_{c_{{\ast},{\star}}}$ represents the registers $c_{\ast}$ and $c_{\star}$.
\par (S1.5) Performing the operation $\tilde{\mathcal{O}}$ on registers $c_{4}$ and $c_{6}$ to load the input image information, when the register $c_{5}$ in state $|0\rangle$. Simultaneously, the data of the convolutional kernel is encoded to the quantum system by applying similar operations on the registers $c_{2}$, $c_{7}$. Consequently, the transformed the quantum state is given as
     \begin{equation}
     \frac{1}{\sqrt{2}{M^{\prime}}N}{\sum\limits_{{m^{\prime}} = 0}^{{M^{\prime}}^{2} - 1}} |  m^{\prime} \rangle _{c_{1}} \sum\limits_{n = 0}^{N^{2} - 1} | {n} \rangle _{c_{2}} |m \rangle_{c_4} ( |0\rangle |\tau_{mn} \rangle + |1\rangle |00\rangle )_{c_{5,6,7}},
     \label{eq:11}
     \end{equation}
where $|\tau_{mn} \rangle = {\tilde{r}}_{m} {\tilde{w}}_{n} |00\rangle + \sqrt{1-\left| {\tilde{r}}_{m} {\tilde{w}}_{n} \right|^{2}} |00\rangle^{\bot}$. The quantum state $| \cdot \rangle^{\bot}$ is denoted as the orthogonal state of $|\cdot \rangle$. Specifically, $|00\rangle^{\bot} = 1/\sqrt{3} (|01\rangle + |10\rangle +|11\rangle)$ in Eq. (\ref{eq:11}). It should be noted that the loaded angle values differ for the operation $\tilde{\mathcal{O}}$ executed twice. In the first execution, the needed angle is $ \vartheta_{m} = \arccos{{\tilde{r}}_{m}}$, as defined in Sec. \ref{sec:3.0}. In the second execution, the required angle is $ \vartheta_{n}^{\prime} = \arccos{{\tilde{w}}_{n}}$ (${{\tilde{w}}_{n}} = {w_{n} / {\left\| W \right\|_{F}}}$ represents the convolution kernel information after data preprocessing).

\par (S1.6) Applying a $Hadamard$ gate on the register $c_{5}$, the quantum system becomes into
     \begin{equation}
     \begin{split}
     \frac{1}{2{M^{\prime}}N}&{\sum\limits_{{m^{\prime}} = 0}^{{M^{\prime}}^{2} - 1}} |  m^{\prime} \rangle _{c_{1}} \sum\limits_{n = 0}^{N^{2} - 1} | {n} \rangle _{c_{2}} |m \rangle_{c_4} \otimes\\
     &\left[ |0\rangle (|\tau_{mn} \rangle + |00\rangle )+ |1\rangle ( |\tau_{mn} \rangle - |00\rangle ) \right]_{c_{5,6,7}}.
     \end{split}
     \label{eq:12}
     \end{equation}
It can be rewritten as
     \begin{equation}
     \begin{split}
     &\frac{1}{M^{\prime}} \sum\limits_{m^{\prime}=0}^{{M^{\prime}}^{2} - 1} |  m^{\prime} \rangle _{c_{1}} |\psi_{m^{\prime}} \rangle_{c_{5,2,4,6,7}} \\
     =&\frac{1}{M^{\prime}} \sum\limits_{m^{\prime}=0}^{{M^{\prime}}^{2} - 1} |  m^{\prime} \rangle _{c_{1}}\left(\sin \theta_{m^{\prime}} |\phi_{m^{\prime}}^{0} \rangle + \cos \theta_{m^{\prime}} |\phi_{m^{\prime}}^{1} \rangle \right)_{c_{5,2,4,6,7}},
     \end{split}
     \label{eq:13}
     \end{equation}
where $|\phi_{m^{\prime}}^{0} \rangle = \frac{|0\rangle \sum_{n=0}^{N^2-1} |n\rangle |m\rangle (|\tau_{mn} \rangle + |00\rangle)}{\sqrt{2(N^{2}+\sum_{n=0}^{N^2-1} {{\tilde{r}}_{m}} {{\tilde{w}}_{n}})}}$, $|\phi_{m^{\prime}}^{1} \rangle = \frac{ |1\rangle \sum_{n=0}^{N^2-1} |n\rangle |m\rangle  (|\tau_{mn} \rangle - |00\rangle)}{\sqrt{2(N^{2}-\sum_{n=0}^{N^2-1} {{\tilde{r}}_{m}} {{\tilde{w}}_{n}})}}$, $\sin{\theta_{m^{\prime}}} = \sqrt{ \frac{N^{2}+\sum_{n=0}^{N^2-1} {{\tilde{r}}_{m}} {{\tilde{w}}_{n}}}{2N^{2}} }$, and $\cos{\theta_{m^{\prime}}} = \sqrt{ \frac{N^{2}-\sum_{n=0}^{N^2-1} {{\tilde{r}}_{m}} {{\tilde{w}}_{n}}}{2N^{2}} }$.
\par In fact, $(2N^{2} \sin^{2}{\theta_{m^{\prime}}} - N^2)$ is the feature extracted by a convolution and stored at position $m^{\prime}$ (i.e., $(x^{\prime}, y^{\prime})$) in the feature map, denoted as $r^{\prime}_{m^{\prime}}$. For each specified $|m^{\prime} \rangle$, all operations to obtain the quantum state $|\psi_{m^{\prime}} \rangle$ are denoted as $\mathcal{A}_{m^{\prime}}$. This means that the unitary operation $\mathcal{A}_{m^{\prime}}$ can implement the transformation $\mathcal{A}_{m^{\prime}} |  m^{\prime} \rangle |0\cdots0\rangle \rightarrow  |  m^{\prime} \rangle |\psi_{m^{\prime}} \rangle$.
\par From Lemma \ref{lem:QAE}, we can construct the unitary operation $Q_{m^{\prime}} = - \mathcal{A}_{m^{\prime}} S_{1} \mathcal{A}_{m^{\prime}}^{\dagger} S_{0}$. The difference is that here $S_{1} = I^{{\otimes qub} + 3} - 2\left( |  0 \rangle \langle 0  | \right)^{\otimes qub + 3}$ and $S_{0} = I^{\otimes qub+1}\otimes\left( I^{\otimes 2} - 2| 00 \rangle \langle 00 | \right)$, where $qub = {6{\lceil{\log M}\rceil}+L+{\lceil{\log s}\rceil}}$. Mathematically, each $|\psi_{m^{\prime}} \rangle$ (as shown in Eq. (\ref{eq:13})) can be Schmidt decomposed into the eigenstates $| \Psi_{m^{\prime}}^{\pm} \rangle = \frac{1}{\sqrt{2}} \left(|\phi_{m^{\prime}}^{0} \rangle \pm \mathbf{i}|\phi_{m^{\prime}}^{1} \rangle  \right)$ ($ \mathbf{i} = \sqrt{-1}$) of $Q_{m^{\prime}}$. These eigenstates correspond to the eigenvalue $e^{\pm 2\mathbf{i}\theta_{m^{\prime}}}$ respectively.
\par (S1.7) A register $c_{8}$ in state $|0\rangle^{\otimes \log\epsilon^{- 1}}$ is appended. Subsequently, the process of QAE \cite{brassard2002} is introduced to generate the state
     \begin{equation}
     \begin{split}
        \frac{1}{M^{\prime}}\sum\limits_{m^{\prime} = 0}^{{M^{\prime}}^{2} - 1} |  m^{\prime} \rangle_{c_{1}} & \frac{-\mathbf{i}}{\sqrt{2}} \left( e^{\mathbf{i}\theta_{m^{\prime}}}| \Psi_{m^{\prime}}^{+} \rangle |  {\tilde{\theta}}_{m^{\prime}} \rangle \right.\\
         & \left.- e^{- \mathbf{i}\theta_{m^{\prime}}} |  \Psi_{m^{\prime}}^{-} \rangle |  {- {\tilde{\theta}}_{m^{\prime}}} \rangle  \right)_{c_{5,2,4,6,7,8}},
     \end{split}
     \label{eq:14}
     \end{equation}
where $| {{\theta_{m^{\prime}}}/{\pi} - {\tilde{\theta}}_{m^{\prime}}} | \leq \epsilon$.

\par (S1.8) According to the fact $r^{\prime}_{m^{\prime}} = 2N^{2} \sin^{2}({\pi \tilde{\theta}_{m^{\prime}}}) - N^2$, an additional register $c_{9}$ is introduced and a unitary operation $U_{f(x)}$ is employed. $U_{f(x)}$ can be implemented a function $f(x) = 2N^{2} \sin^{2}(x) - N^2$ follow Lemma \ref{lem:QAC}. Then, we can get the state
     \begin{equation}
     \frac{1}{M^{\prime}} \sum\limits_{m^{\prime}=0}^{{M^{\prime}}^{2} - 1} |  m^{\prime} \rangle _{c_{1}} |\psi_{m^{\prime}} \rangle_{c_{5,2,4,6,7,8}}|r^{\prime}_{m^{\prime}}\rangle_{c_{9}}.
     \label{eq:15}
     \end{equation}
\par
\begin{figure}[htbp!]
        \centering
        \includegraphics[height=0.15\textwidth,width=0.48\textwidth]{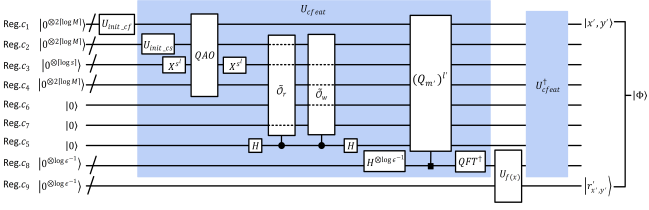}
        \caption{The circuit for the quantum convolution layer in QCNNFS model. Here, $U_{init\_cf}$, $U_{init\_cs}$ mark the operations of initializing the quantum system in step (S1.0), i.e., $U_{init\_cf} = I^{\otimes( 2{\lceil{\log{M}}\rceil} - 2{\lceil{\log{M^{\prime}}}\rceil} )} \otimes H^{ \otimes 2{\lceil{\log{M^{\prime}}}\rceil} }$ and $U_{init\_cs} = I^{\otimes( 2{\lceil{\log{M}}\rceil} - 2{\lceil{\log{N}}\rceil} )} \otimes H^{ \otimes 2{\lceil{\log{N} }\rceil} }$ ($H$ is represented as the $Hadamard$ gate). $\tilde{\mathcal{O}_{r}}$, $\tilde{\mathcal{O}_{w}}$ are labeled as the operation $\tilde{\mathcal{O}}$, with different subscripts to distinguish the loaded information. And, the variables $l =1, \cdots, {\lceil \log{s} \rceil}$, $l^{\prime} =2^{0}, \cdots, 2^{\log\epsilon^{- 1}-1}$. Moreover, the dashed line indicates that the qubits do not undergo the current unitary gate.}
    \label{fig:Ucon}
    \end{figure}

 (S1.9) To reuse the intermediate registers in successive layers, the inverse operations of QAE and $\mathcal{A}_{m^{\prime}}$ are implemented. At this point, the quantum state
     \begin{equation}
     \begin{split}
     |\Phi\rangle &= \frac{1}{M^{\prime}} \sum\limits_{m^{\prime}=0}^{{M^{\prime}}^{2} - 1} |  m^{\prime} \rangle _{c_{1}} |r^{\prime}_{m^{\prime}}\rangle_{c_{9}}\\
     &= \frac{1}{M^{\prime}} \sum\limits_{m^{\prime}=0}^{{M^{\prime}} - 1} |  x^{\prime}, y^{\prime} \rangle _{c_{1}} |r^{\prime}_{x^{\prime}, y^{\prime}}\rangle_{c_{9}}
     \end{split}
     \label{eq:16}
     \end{equation}
can be obtained. This quantum state represents the feature information of the image after convolution, such as the feature information at position $({x^{\prime}, y^{\prime}})$ is $r^{\prime}_{x^{\prime}, y^{\prime}}$. Moreover, $|\Phi\rangle$ can be regarded as the input of the next layer. The quantum circuit to implemented the convolution layer is demonstrated in Fig. \ref{fig:Ucon}.
\subsection{\label{sec:3.2}Quantum pooling layer}
\par A method for implementing pooling layers in the quantum system is proposed to enhance the robustness of the model, analogous to classical CNN \cite{jiao202036}. We focus on the case of average pooling, which could smooth feature maps and reduce overfitting \cite{lin2013network}. Assuming that the pooling layer uses a sliding window with size $N^{\prime} \times N^{\prime}$ and a preset stride $s^{\prime}$ $(s^{\prime}>0,s^{\prime}\in \mathbb{Z})$. Additionally, the input to the pooling layer can be the output quantum state $|\Phi_{P}\rangle$ from any preceding layer. This quantum state may comes from either the production of a quantum convolution layer or a quantum pooling layer. Input information can also be loaded from QRAM using the method described in Sec. \ref{sec:3.1}. For simplicity, let the state $|\Phi\rangle$ in Eq. (\ref{eq:16}) be used as the input to this layer, i.e., $|\Phi_{P}\rangle = |\Phi\rangle$.
\par
\begin{figure}
        \centering
        \includegraphics[height=0.15\textwidth, width=0.45\textwidth]{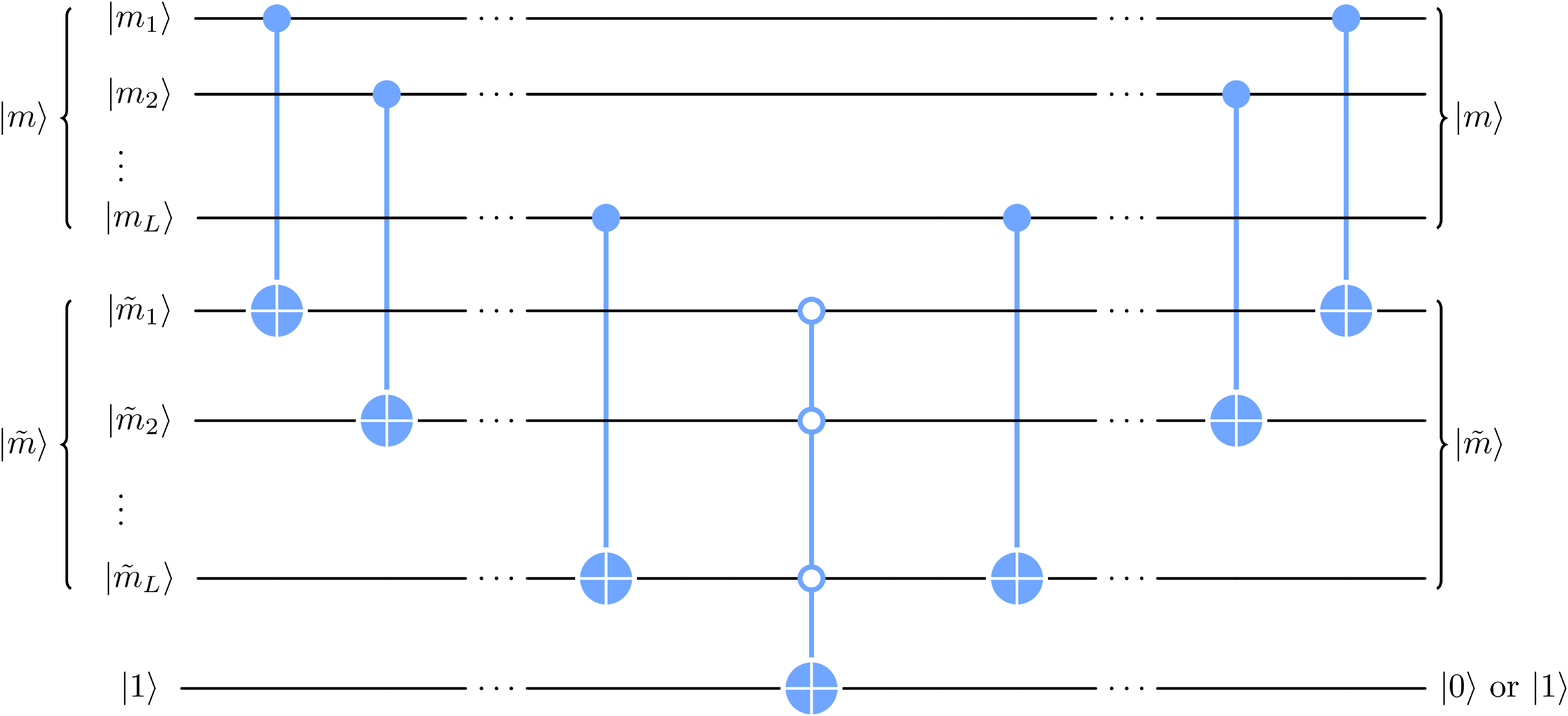}
        \caption{The quantum circuit to implement the unitary gate $U_{c}$.}
    \label{fig:Uc}
    \end{figure}

 Before introducing the quantum pooling layer, we design a unitary gate
    \begin{equation}
        {U}_{c}|m\rangle |\tilde{m}\rangle |1\rangle =
        \begin{cases}
            |m\rangle |\tilde{m}\rangle |0\rangle,&m = \tilde{m} \\
            |m\rangle |\tilde{m}\rangle |1\rangle, &m \neq \tilde{m}
        \end{cases},
     \label{eq:17}
     \end{equation}
that facilitates information exchange between the preceding and current layers without relying on QRAM. Here, quantum states $|m\rangle$ and $| \tilde{m} \rangle$ are represented by binary sequences of equal length, which are utilized to depict different quantum states. The corresponding quantum circuit for ${U}_{c}$ is shown in Fig. \ref{fig:Uc}. In this context, the implementation processes of quantum pooling are described as follows.
\par (S2.0) According to the pooling hyperparameters (sliding window size $N^{\prime} \times N^{\prime}$ and stride $s^{\prime}$), the indexing state
     \begin{equation}
        \frac{1}{M^{\prime \prime} N^{\prime}} \sum_{x^{\prime \prime}, y^{\prime \prime}=0}^{{M^{\prime \prime}} - 1} |x^{\prime \prime}, y^{\prime \prime} \rangle_{p_{1}} \sum_{i^{\prime} , j^{\prime} = 0}^{{N^{\prime}}-1} |i^{\prime} , j^{\prime} \rangle_{p_{2}} |s^{\prime} \rangle_{p_{3}}
     \label{eq:18}
     \end{equation}
is prepared. Among them, ${M^{\prime \prime}} = \frac{M^{\prime} - N^{\prime}}{s^{\prime}} + 1$ is labeled as the edge length of the feature map extracted by the pooling layer, $|x^{\prime \prime}, y^{\prime \prime}\rangle$ is the quantum state encoding the positional information of features after pooling, and $|i^{\prime} , j^{\prime} \rangle$ is represented as the positional information of the coverage region of the sliding window.
\par (S2.1) By performing similar operations as steps (S1.2)-(S1.3) in Sec. \ref{sec:3.1}, we can get the state
     \begin{equation}
        \frac{1}{M^{\prime \prime} N^{\prime}} \sum_{x^{\prime \prime}, y^{\prime \prime}=0}^{{M^{\prime \prime}} - 1} |x^{\prime \prime}, y^{\prime \prime} \rangle_{p_{1}} \sum_{i^{\prime} , j^{\prime} = 0}^{{N^{\prime}}-1} |i^{\prime} , j^{\prime} \rangle_{p_{2}} |\tilde{x}^{\prime}, \tilde{y}^{\prime} \rangle_{p_{4}},
     \label{eq:19}
     \end{equation}
where $\tilde{x}^{\prime} = x^{\prime \prime}s^{\prime} + i^{\prime}$, $\tilde{y}^{\prime} = y^{\prime \prime}s^{\prime} + j^{\prime}$ are denoted as the positional information of the features involved in the sliding window. We can also rewrite $|m^{\prime \prime}\rangle = |x^{\prime \prime}, y^{\prime \prime}\rangle$, $|n^{\prime}\rangle = |i^{\prime} , j^{\prime} \rangle$, and $|\tilde{m}^{ \prime}\rangle = |\tilde{x}^{\prime}, \tilde{y}^{\prime}\rangle$. At this point, the state of the entire quantum system can be described as
     \begin{equation}
        \begin{split}
            &|\Phi\rangle \otimes \frac{1}{M^{\prime \prime} N^{\prime}} \sum\limits_{m^{\prime \prime}=0}^{{M^{\prime \prime}}^{2} - 1} |{m^{\prime \prime}}\rangle_{p_{1}} \sum\limits_{n^{\prime} = 0}^{{N^{\prime}}^{2}-1} |n^{\prime} \rangle_{p_{2}} | \tilde{m}^{ \prime}\rangle_{p_{4}}\\
            =&\frac{1}{M^{\prime}} \sum\limits_{m^{\prime}=0}^{{M^{\prime}}^{2} - 1} |  m^{\prime} \rangle _{c_{1}} |r^{\prime}_{m^{\prime}}\rangle_{c_{9}} \otimes\\
             &\frac{1}{M^{\prime \prime} N^{\prime}} \sum\limits_{m^{\prime \prime}=0}^{{M^{\prime \prime}}^{2} - 1} |{m^{\prime \prime}}\rangle_{p_{1}} \sum\limits_{n^{\prime} = 0}^{{N^{\prime}}^{2}-1} |n^{\prime} \rangle_{p_{2}} | \tilde{m}^{ \prime}\rangle_{p_{4}}.
        \end{split}
     \label{eq:20}
     \end{equation}
\par (S2.2) We append a register $p_{5}$ in state $|1\rangle$. The unitary gate $U_{c}$ is implemented on the registers $c_{1}$, $p_{4}$, and $p_{5}$, the quantum state
     \begin{equation}
        \begin{split}
            &\frac{1}{M^{\prime} M^{\prime \prime} N^{\prime}}
              \sum\limits_{m^{\prime \prime}=0}^{{M^{\prime \prime}}^{2} - 1} |{m^{\prime \prime}}\rangle_{p_{1}} \sum\limits_{n^{\prime} = 0}^{{N^{\prime}}^{2}-1} |n^{\prime} \rangle_{p_{2}} | \tilde{m}^{ \prime}\rangle_{p_{4}} \otimes \\
              &\Bigl( |m^{\prime}\rangle_{c_{1}} |r^{\prime}_{m^{\prime}}\rangle_{c_{9}} |0\rangle_{p_{5}} + \sum\limits_{m^{\prime}\neq \tilde{m}^{\prime}}^{{M^{\prime}}^{2} - 1} |  m^{\prime} \rangle _{c_{1}} |r^{\prime}_{m^{\prime}}\rangle_{c_{9}}|1\rangle_{p_{5}} \Bigr)
        \end{split}
     \label{eq:21}
     \end{equation}
is obtained.
\par (S2.3) Two registers $p_{6}$, $p_{7}$ are added, each initialized to the state $|0\rangle$. A $Hadamard$ gate is applied on the register $p_{6}$ when the register $p_{5}$ is in the state $|0\rangle$. Subsequently, if registers $p_{5}$ and $p_{6}$ in the state $|00\rangle$, a controlled rotation \cite{congLDA2016} is performed on the register $p_{7}$ to generate
     \begin{equation}
        \begin{split}
            &\frac{1}{ M^{\prime} M^{\prime \prime} N^{\prime}}
              \sum\limits_{m^{\prime \prime}=0}^{{M^{\prime \prime}}^{2} - 1} |{m^{\prime \prime}}\rangle_{p_{1}} \sum\limits_{n^{\prime} = 0}^{{N^{\prime}}^{2}-1} |n^{\prime} \rangle_{p_{2}} | \tilde{m}^{ \prime}\rangle_{p_{4}} \otimes \\
              &\Bigl[ \frac{1}{\sqrt{2}}|\tilde{m}^{\prime}\rangle_{c_{1}} |r^{\prime}_{\tilde{m}^{\prime}}\rangle_{c_{9}} \left(|00\rangle_{p_{5,6}} |\tau_{\tilde{m}^{\prime}n^{\prime}}\rangle_{p_{7}} +|01\rangle_{p_{5,6}} |0\rangle_{p_{7}}\right) \\
              & + \sum\limits_{m^{\prime}\neq \tilde{m}^{\prime}}^{{M^{\prime}}^{2} - 1} |  m^{\prime} \rangle _{c_{1}} |r^{\prime}_{m^{\prime}}\rangle_{c_{9}} |10\rangle_{p_{5,6}} |0\rangle_{p_{7}}\Bigr],
        \end{split}
     \label{eq:22}
     \end{equation}
where $|\tau_{\tilde{m}^{\prime}n^{\prime}} \rangle = {r^{\prime}}_{\tilde{m}^{\prime}} |0\rangle + \sqrt{1-\left| {r^{\prime}}_{\tilde{m}^{\prime}} \right|^{2}} |1\rangle$. Here, the control qubits for the controlled rotation operation are corresponding the qubits in the register $c_{9}$.
\par (S2.4) Another $Hadamard$ gate is implemented on the register $p_{6}$ when the register $p_{5}$ in state $|0\rangle$. The system state becomes into
     \begin{equation}
        \frac{1}{M^{\prime \prime}}
              \sum\limits_{m^{\prime \prime}=0}^{{M^{\prime \prime}}^{2} - 1} |{m^{\prime \prime}}\rangle_{p_{1}} (\sin \theta_{m^{\prime\prime}}^{\prime} |\varphi_{m^{\prime\prime}}\rangle + \cos \theta_{m^{\prime\prime}}^{\prime} |\varphi_{m^{\prime\prime}}^{\bot}\rangle )_{p_{2,4},c_{1,9},p_{5,6,7}},
     \label{eq:23}
     \end{equation}
where
     \begin{equation}
        \begin{split}
            &|\varphi_{m^{\prime\prime}}\rangle = \frac{1}{2 M^{\prime} N^{\prime} \sin \theta_{m^{\prime\prime}}^{\prime}} \sum\limits_{n^{\prime} = 0}^{{N^{\prime}}^{2}-1} |n^{\prime} \rangle_{p_{2}} | \tilde{m}^{ \prime}\rangle_{p_{4}} |\tilde{m}^{\prime}\rangle_{c_{1}} \otimes \\
            &|r^{\prime}_{\tilde{m}^{\prime}}\rangle_{c_{9}} |00\rangle_{p_{5,6}}\left( |\tau_{\tilde{m}^{\prime}n^{\prime}}\rangle + |0\rangle \right)_{p_{7}},
        \end{split}
     \label{eq:23+1}
     \end{equation}
and
     \begin{equation}
        \begin{split}
            &|\varphi_{m^{\prime\prime}}^{\bot}\rangle = \frac{1}{ 2M^{\prime} N^{\prime} \cos \theta_{m^{\prime\prime}}^{\prime}} \sum\limits_{n^{\prime} = 0}^{{N^{\prime}}^{2}-1} |n^{\prime} \rangle_{p_{2}} | \tilde{m}^{ \prime}\rangle_{p_{4}} \otimes \\
            &\Bigl[|\tilde{m}^{\prime}\rangle_{c_{1}} |r^{\prime}_{\tilde{m}^{\prime}}\rangle_{c_{9}}|01\rangle_{p_{5,6}}\left( |\tau_{\tilde{m}^{\prime}n^{\prime}}\rangle - |0\rangle \right)_{p_{7}} + \\
            & 2\sum\limits_{m^{\prime}\neq \tilde{m}^{\prime}}^{{M^{\prime}}^{2} - 1} |  m^{\prime} \rangle _{c_{1}} |r^{\prime}_{m^{\prime}}\rangle_{c_{9}} |10\rangle_{p_{5,6}} |0\rangle_{p_{7}}\Bigr].
        \end{split}
     \label{eq:23+2}
     \end{equation}
We denote $\sin \theta_{m^{\prime\prime}}^{\prime} = ({{N^{\prime}}^{2} + \sum_{n^{\prime} = 0}^{{N^{\prime}}^{2}-1} {{{r^{\prime}}}_{\tilde{m}^{\prime}}} })^{\frac{1}{2}} / {\sqrt{2} M^{\prime} N^{\prime}}$, $\cos \theta_{m^{\prime\prime}}^{\prime} = ({2{M^{\prime}}^{2}{N^{\prime}}^{2}-{N^{\prime}}^{2}- \sum_{n^{\prime} = 0}^{{N^{\prime}}^{2}-1} {{{r^{\prime}}}_{\tilde{m}^{\prime}}}})^{\frac{1}{2}} /{\sqrt{2} M^{\prime} N^{\prime}}$.
\par (S2.5) According to Lemma \ref{lem:QAE}, by performing operations similar to steps (S1.7)-(S1.8) in Sec. \ref{sec:3.1}, we can get
     \begin{equation}
        \begin{split}
            &\frac{1}{M^{\prime \prime}}
              \sum\limits_{m^{\prime \prime}=0}^{{M^{\prime \prime}}^{2} - 1} |{m^{\prime \prime}}\rangle_{p_{1}} \left(\sin \theta_{m^{\prime\prime}}^{\prime} |\varphi_{m^{\prime\prime}}\rangle |\theta_{m^{\prime\prime}}^{\prime}\rangle + \right.\\
              &\left. \cos \theta_{m^{\prime\prime}}^{\prime} |\varphi_{m^{\prime\prime}}^{\bot}\rangle |-\theta_{m^{\prime\prime}}^{\prime}\rangle \right)_{p_{2,4},c_{1,9},p_{5,6,7}} |r^{\prime \prime}_{m^{\prime \prime}}\rangle_{p_{9}},
        \end{split}
     \label{eq:24}
     \end{equation}
where $r^{\prime \prime}_{m^{\prime \prime}} = 2{M^{\prime}}^{2}{N^{\prime}}^{2} \sin^{2}\theta_{m^{\prime\prime}}^{\prime} -{N^{\prime}}^{2}$ is represented as a feature at position $(x^{\prime \prime},y^{\prime \prime})$ in the feature map obtained after the pooling layer.
\par (S2.6) Performing the inverse operations of steps (S2.3-S2.5), the entire system transitions into the state
     \begin{equation}
            |\Phi\rangle_{c_{1,9}} \otimes \frac{1}{M^{\prime \prime}}
              \sum\limits_{m^{\prime \prime}=0}^{{M^{\prime \prime}}^{2} - 1} |{m^{\prime \prime}}\rangle_{p_{1}} |r^{\prime \prime}_{m^{\prime \prime}}\rangle_{p_{9}}.
     \label{eq:25}
     \end{equation}
To reuse intermediate registers across successive layers, the inverse operations of the quantum convolution are applied, we can obtain the state
     \begin{equation}
        \begin{split}
           |\Phi^{\prime}\rangle_{p_{1,9}} = &\frac{1}{M^{\prime \prime}}
              \sum\limits_{m^{\prime \prime}=0}^{{M^{\prime \prime}}^{2} - 1} |{m^{\prime \prime}}\rangle_{p_{1}} |r^{\prime \prime}_{m^{\prime \prime}}\rangle_{p_{9}}\\
              =&\frac{1}{M^{\prime \prime}}
              \sum\limits_{x^{\prime \prime}=0, y^{\prime \prime}=0}^{{M^{\prime \prime}} - 1} |x^{\prime \prime},y^{\prime \prime}\rangle_{p_{1}} |r^{\prime \prime}_{x^{\prime \prime},y^{\prime \prime}}\rangle_{p_{9}}
        \end{split}
     \label{eq:26}
     \end{equation}
with the registers $p_{1}$, $p_{9}$. It encapsulates the feature information of the image after pooling and can serve as input for the subsequent quantum algorithm.
    \begin{figure}[htbp!]
        \centering
        \includegraphics[width=0.48\textwidth]{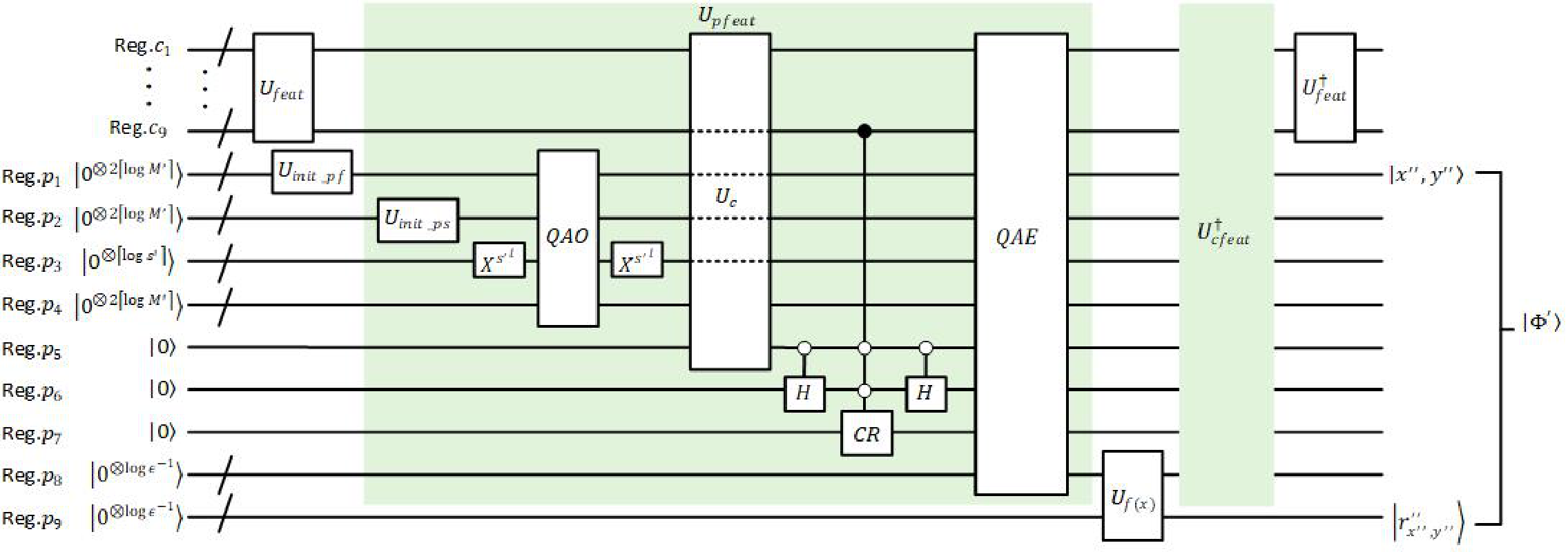}
        \caption{The circuit for the quantum pooling layer. In the diagram, $U_{feat}$ is marked as the operations to prepare the input state $|\Phi\rangle$ of this pooling layer. $U_{init\_cf} = I^{\otimes( 2{\lceil{\log{M^{\prime}}}\rceil} - 2{\lceil{\log{M^{\prime\prime}}}\rceil} )} \otimes H^{ \otimes 2{\lceil{\log{M^{\prime\prime}}}\rceil} }$ and $U_{init\_cs} = I^{\otimes( 2{\lceil{\log{M^{\prime}}}\rceil} - 2{\lceil{\log{N^{\prime}}}\rceil} )} \otimes H^{ \otimes 2{\lceil{\log{N^{\prime}} }\rceil} }$ are represented as the operation of initializing the index information in Eq. (\ref{eq:18}). Here, the value of $l$ is $1, \cdots, {\lceil \log{s^{\prime}} \rceil}$.}
    \label{fig:Upool}
    \end{figure}
\subsection{\label{sec:3.3}Quantum fully connected layer}
\par Corresponding to classical CNN, an implementation method for the quantum fully connected layer is discussed. It can directly process the quantum state output by quantum convolution layers or quantum pooling layers and predict the classification result of the image.
\par Simply, we discuss a fully connected layer with $K$ outputs for the classification of $K$ classes. Supposing that the quantum state of the image information processed by a certain number of quantum convolution and quantum pooling operations is denoted as $|\Phi_{F}\rangle = \frac{1}{\bar{M}}\sum_{\bar{x}=0, \bar{y}=0}^{\bar{M} - 1} |\bar{x}, \bar{y}\rangle_{f_{1}} |\bar{r}_{\bar{x}, \bar{y}}\rangle_{f_{2}}$, where $\bar{M}$ is the number of features sufficient to identify the image, $\bar{r}_{\bar{x}, \bar{y}}$ is the pixel feature located at position $(\bar{x}, \bar{y})$. Furthermore, this layer has $K$ sets of weight vectors already stored in QRAM, denoted as $\{ w_{\bar{x}, \bar{y}}^{k} \}_{\bar{x}, \bar{y} = 0}^{\bar{M} - 1}$ $(k=0,1,\cdots,K-1)$. In this case, the quantum fully connected layer is implemented as shown in the following.
\par (S3.0) We first prepare the state $\frac{1}{\sqrt{K}} \sum_{k=0}^{K-1} |k\rangle$. Then, the quantum system state can be described as $\frac{1}{\bar{M}\sqrt{K}} \sum_{\bar{x}=0, \bar{y}=0}^{\bar{M} - 1} \sum_{k=0}^{K-1} |\bar{x}, \bar{y}\rangle_{f_{1}} |\bar{r}_{\bar{x}, \bar{y}}\rangle_{f_{2}} |k\rangle_{f_{3}}$.
\par (S3.1) Adding three registers is in the state $|000\rangle_{f_{4,5,6}}$. A $Hadamard$ gate is applied on the register $f_{6}$ to generate the state
     \begin{equation}
        \begin{split}
            &\frac{1}{\bar{M}\sqrt{K}} \sum_{\bar{x}=0, \bar{y}=0}^{\bar{M} - 1} \sum_{k=0}^{K-1} |\bar{x}, \bar{y}\rangle_{f_{1}} |\bar{r}_{\bar{x}, \bar{y}}\rangle_{f_{2}} |k\rangle_{f_{3}} \otimes \\
            &|00\rangle_{f_{4,5}}\frac{1}{\sqrt{2}}(|0\rangle + |1\rangle)_{f_{6}}.
        \end{split}
     \label{eq:27}
     \end{equation}
\par (S3.2) When the register $f_{6}$ in the state $|0\rangle$, the register $f_{3}$ serves as the control qubits. The operation $\tilde{\mathcal{O}}$ is controlled by the register $f_{3}$ and performed on the registers $f_{1,4}$ to encode the $k$-th weight information into the quantum system. Then, the state
     \begin{equation}
        \begin{split}
            &\frac{1}{\bar{M}\sqrt{2K}} \sum_{\bar{x}=0, \bar{y}=0}^{\bar{M} - 1} \sum_{k=0}^{K-1} |\bar{x}, \bar{y}\rangle_{f_{1}} |\bar{r}_{\bar{x}, \bar{y}}\rangle_{f_{2}} |k\rangle_{f_{3}} \otimes \\
            &[(w_{\bar{x}, \bar{y}}^{k}|0\rangle + \sqrt{1-|w_{\bar{x}, \bar{y}}^{k}|^{2}} |1\rangle)|00\rangle + |001\rangle]_{f_{4,5,6}}
        \end{split}
     \label{eq:28}
     \end{equation}
can be obtained.
\par (S3.3) Follow the way in step (S2.3), if the register $f_{6}$ in the state $|0\rangle$, a controlled rotation is implemented on the register $f_{5}$, with the control determined by the state $|\bar{r}_{\bar{x}, \bar{y}}\rangle_{f_{2}}$. The system then transforms into the state
     \begin{equation}
        \begin{split}
            &\frac{1}{\bar{M}\sqrt{2K}} \sum_{\bar{x}=0, \bar{y}=0}^{\bar{M} - 1} \sum_{k=0}^{K-1} |\bar{x}, \bar{y}\rangle_{f_{1}} |\bar{r}_{\bar{x}, \bar{y}}\rangle_{f_{2}} |k\rangle_{f_{3}} \otimes \\
             &(|\tau_{\bar{x}, \bar{y}}^{k} \rangle_{f_{4,5}} |0\rangle_{f_{6}} +
            |00\rangle_{f_{4,5}} |1\rangle_{f_{6}}),
        \end{split}
     \label{eq:29}
     \end{equation}
where $|\tau_{\bar{x}, \bar{y}}^{k} \rangle = \bar{r}_{\bar{x}, \bar{y}} w_{\bar{x}, \bar{y}}^{k} |00\rangle + \sqrt{1- |\bar{r}_{\bar{x}, \bar{y}} w_{\bar{x}, \bar{y}}^{k}|^2} |00\rangle^{\bot}$.
\par (S3.4) A $Hadamard$ gate is applied on the register $f_{6}$ to produce
     \begin{equation}
        \begin{split}
            &\frac{1}{\bar{M}\sqrt{2K}} \sum_{\bar{x}=0, \bar{y}=0}^{\bar{M} - 1} \sum_{k=0}^{K-1} |\bar{x}, \bar{y}\rangle_{f_{1}} |\bar{r}_{\bar{x}, \bar{y}}\rangle_{f_{2}} |k\rangle_{f_{3}} \otimes \\
             &[(|\tau_{\bar{x}, \bar{y}}^{k} \rangle + |00\rangle )_{f_{4,5}} |0\rangle_{f_{6}} +
            (|\tau_{\bar{x}, \bar{y}}^{k} \rangle - |00\rangle )_{f_{4,5}} |1\rangle_{f_{6}})].
        \end{split}
     \label{eq:30}
     \end{equation}
\par (S3.5) Finally, we measure the registers $f_{3,6}$ in a basis $\{ |k\rangle|0\rangle, |k\rangle|1\rangle\}_{k=0}^{K-1}$, the output probability distribution of $|k\rangle|0\rangle$ is obtained as
     \begin{equation}
        Prob(k,0) = \frac{1}{2K{\bar{M}}^2} \Bigl({\bar{M}}^2 + \sum_{\bar{x}, \bar{y} = 0}^{\bar{M} - 1}\bar{r}_{\bar{x}, \bar{y}} w_{\bar{x}, \bar{y}}^{k} \Bigr).
     \label{eq:31}
     \end{equation}
After performing $O(K)$ measurements, the category $k$ with the highest probability $Prob(k,0)$ can be selected as the predicted category of the image. This yields the output of the fully connected layer neural network and completes the classification task of the CNN.
\section{\label{sec:4}Analysis}
\par In this section, we analyze the resources of the proposed method. In Sec. \ref{sec:4.1}, the time complexity is presented, and the memory complexity for QCNNFS is discussed in Sec. \ref{sec:4.2}.
\subsection{\label{sec:4.1}Time complexity analysis}
\par As previously mentioned, the QCNNFS primarily consists of quantum convolutional layer, quantum pooling layer, and quantum fully connected layer. In this section, we will analyze the time complexity of each layer individually.
\par \textbf{1) The time complexity of quantum convolution layer.} In the process of using quantum circuit to realize the convolution operation, the first step is to initialize a quantum system in time $O(\log(M^{\prime}N) )$. In step (S1.1), the operations $X^{s^{l^{\prime}}}$ $(l^{\prime}=1, \cdots, {\lceil \log{s} \rceil})$ are applied to preparing the state $|s\rangle$. Generally, the stride $s$ is a small integer, thus its time complexity is $O(1)$. Follow Lemma \ref{lem:QAC}, step (S1.2) is completed in time $O(\text{poly} \log M)$. The inverse operations of step (S1.1) are performed in step (S1.3). It takes $O(1)$ time. A $Hadamard$ gate is executed in step (S1.4) which requires time $O(1)$. In step (S1.5), $\tilde{\mathcal{O}}$ is used to load the data information in time $O(\text{poly} \log (M^{2}N^{2}) + L)$, according to the Sec. \ref{sec:3.0}. The time complexity of step (S1.6) is $O(1)$, similar to step (S1.4). In step (S1.7), a unitary operation $Q_{m^{\prime}}$ is implemented in time $O(\log(M^{\prime}N) + \text{poly} \log (M^{2}N^{2}) + L)$, which includes the operations of step (S1.0)-(S1.6). And this step needs $\epsilon^{-1}$ copies of $Q_{m^{\prime}}$ to estimated $\theta_{m^{\prime}}$ within the error $\epsilon$. Hence, the time complexity of step (S1.7) is $O[(\log(M^{\prime}N) + \text{poly} \log (M^{2}N^{2}) + L)\epsilon^{-1}]$. In Step (S1.8), the running time $O(\text{poly}\log{\epsilon^{-1}})$ for realizing the quantum arithmetic operations. It should be noted that the time of steps (S1.2) and (S1.8) can be ignored because they are much smaller than QAE. Step (S1.9) takes time $O[(\log(M^{\prime}N) + \text{poly} \log (M^{2}N^{2}) + L)\epsilon^{-1}]$ to perform the inverse operation of QAE and $\mathcal{A}_{m^{\prime}}$. Therefore, the time required to implement convolutional layer operation with quantum circuit is $O[(\log(M^{\prime}N) + \text{poly} \log (M^{2}N^{2}) + L)\epsilon^{-1}]$.
\par \textbf{2) The time complexity of quantum pooling layer.} Here, we analyze the time complexity of pooling operations in quantum system. First, it is necessary to prepare the input state $|\Phi_{P}\rangle$. Assuming that the time it takes is $O(T_{\Phi_{P}})$. Then, the state in step (S2.0) can be got in time $O(\log{(M^{\prime \prime} N^{\prime})})$. For step (S2.1), the similar operations as step (S1.2)-(S1.3) are implemented in time $O(\text{poly} \log{M^{\prime}})$. Subsequently, ${U}_{c}$ is applied in time $O(\text{poly} \log{M^{\prime}})$, which takes $O(\text{poly} \log{M^{\prime}})$ elementary gates \cite{nielsen2010quantum,barenco1995elementary}. In step (S2.4), the controlled rotation is performed in time $O(\log{\epsilon^{-1}})$ \cite{congLDA2016}. Step (S2.5) needs $O(1)$ time to apply a $Hadamard$ gate. The operations of step (S2.6) is similar to steps (S1.7)-(S1.8), the time complexity is $O\{[\log(M^{\prime \prime}N^{\prime}) + \text{poly} \log {M^{\prime}} + \log{\epsilon^{-1}} + T_{\Phi_{P}}]\epsilon^{-1}\}$.
\par \textbf{3) The time complexity of quantum fully connected layer.} We first consider the input state $|\Phi_{F}\rangle$ is prepared in time $O(T_{\Phi_{F}})$. Step (S3.0) generates a quantum state corresponding to the class, which needs $O(\log{K})$ time. Then, taking a $Hadamard$ gate in time $O(1)$. In step (S3.2), taking $O(\text{poly} \log(\bar{M}K) + L)$ time to encode the weight information. Similar to step (S2.4), step (S3.4) needs $O(\log{\epsilon^{-1}})$ to implement the controlled rotation. Another $H$ gate is used in time $O(1)$, in step (S3.5). Finally, the image's category is determined in step (S3.6). It costs $O(K)$ time to perform $O(K)$ measurements. Hence, the runtime of quantum fully connected layer is $O[K(\text{poly} \log(\bar{M}K) + L + \log{\epsilon^{-1}} + T_{\Phi_{F}})]$.
\par Considering a simple model (shown in Fig. \ref{fig:1b}), which consists a convolution layer, a pooling layer, and a pooling layer. In total, the runtime is $O\{ K[(\log(M^{\prime}N) + \text{poly} \log (M^{2}N^{2}) + L +\log{\epsilon^{-1}})\epsilon^{-2}] \}$. This means that the runtime is $O( K\text{poly} \log (M^{2}N^{2}))$ when $L = \log\epsilon^{- 1} = \lceil\log M\rceil$. Although the time complexity of our algorithm is related to the number of categories $K$, it is often much smaller than the scale of the data. Hence, the proposed algorithm can achieve an exponential acceleration in data scale, compared with the classical counterpart whose runtime is $O(M^2N^2)$. It indicates that our method attains quantum speedup comparable to existing related works \cite{kerenidis2019QCNN,li2020QCNN}, which further embodies the QIP advantages.
\subsection{\label{sec:4.2}Memory complexity analysis}
\par Similar to classical algorithms, memory resources are equally important in QIP. The difference is that QIP relies on qubits as its spatial carrier. Here, we analyze the number of qubits required for the proposed methods.
\par The quantum resources include the storage memory needed for QRAM \cite{kerenidis2017quantum} and the working memory necessary for execution. These resources are primarily related to the scale of the data being processed. In CNN, the image features processed at each layer typically decrease, especially after passing through a pooling layer. For simplicity, we focus on the quantum resources needed for the initial two layers. These layers generally consist of a convolution layer and a pooling layer, which together form a feature extraction combination that reduces the dimensionality of the data \cite{lecun2015deep}. The functions of the steps and the corresponding number of qubits are given in Tab. \ref{tab:table1}.
\par As illustrated in Tab. \ref{tab:table1}, the storage memory of the proposed method is $M^{2} (2\lceil \log M \rceil^{2}+L) + N^{2} (2\lceil \log N \rceil^{2} + L) $ qubits, which is influenced by the amount of data ($M^{2}$ or $N^{2}$) and the storage precision $2^{-L}$. As for the working memory, the qubits in the QCNNFS are mainly utilized for two functions. One is to prepare the index state, which requires preparing three equal-sized registers to perform quantum arithmetic \cite{brassard2002} to get the index of the input information. The other is the feature extraction operation, which initializes a register of $\log\epsilon^{- 1}$ qubits to store information about extracting features. In summary, the working memory of the proposed method is at most $6\lceil \log{M} \rceil + 6\lceil \log{M^{\prime}} \rceil +6 + 2\log\epsilon^{- 1} $. It is important to note that the positional information of the input data can be constructed into the register $c_2$ ($p_2$) in step (S1.3) (step (S2.1)), allowing the omission of register $c_4$ ($p_{4}$). While the quantum convolution process requires using the indices $\{n\}_{n=0}^{N-1}$ in register $c_{2}$ to load the kernel weights is needed in the quantum convolution process, a simple adjustment in the procedure can effectively resolve this concern (i.e., loading the kernel weights first, then constructing the positional information of the data). In this case, the working memory of QCNNFS can be reduced to $4\lceil \log{M} \rceil + 4\lceil \log{M^{\prime}} \rceil +6 + 2\log\epsilon^{- 1} $.
\begin{table*}
\centering
\caption{\label{tab:table1}The functions of steps in the proposed quantum convolution and quantum pooling methods, along with the corresponding number of qubits required.}
\renewcommand\arraystretch{1.5}
\begin{tabular}{m{0.12\textwidth}<{\centering} m{0.06\textwidth}<{\centering} m{0.17\textwidth}<{\centering} m{0.17\textwidth}<{\centering} m{0.008\textwidth}<{\centering} m{0.06\textwidth}<{\centering} m{0.16\textwidth}<{\centering} m{0.18\textwidth}<{\centering}}
\hline
\multirow{2}{*}{Function}  & \multicolumn{3}{c}{Quantum convolution layer} & & \multicolumn{3}{c}{Quantum pooling layer} \\
\cline{2-4} \cline{6-8}
&Steps    &Number of required qubits  &Number of reusable qubits & &Steps  &Number of required qubits  &Number of recoverable qubits\\
\hline
Data storage (QRAM) &- &$M^{2} (2\lceil \log M \rceil^{2}+L) + N^{2} (2\lceil \log N \rceil^{2} + L) $  &- & &- &- &- \\
\hline
Preparing the indexing state  & (S1.0)-(S1.3) & $6\lceil \log{M} \rceil + \lceil \log s \rceil$ & $\lceil \log s \rceil$ & & (S2.0)-(S2.1) & $6\lceil \log{M^{\prime}} \rceil + \lceil \log s^{\prime} \rceil$  &$\lceil \log s^{\prime} \rceil$\\
Loading information    & (S1.4)-(S1.5) & $6\lceil \log{M} \rceil + L + 3$  & $L$ & & (S2.2) & $6\lceil \log{M^{\prime}} \rceil + 1$ &-\\
Extracting features     & (S1.6)-(S1.9) & $6\lceil \log{M} \rceil + 3 + \log\epsilon^{- 1}$ & $4\lceil \log{M} \rceil + 3 + 2(\lceil \log{M} \rceil - \lceil \log{M^{\prime}} \rceil)$ & & (S2.3)-(S2.6) & $6\lceil \log{M} \rceil + 6\lceil \log{M^{\prime}} \rceil +6 + 2\log\epsilon^{- 1}$ &$6\lceil \log{M} \rceil + 4\lceil \log{M^{\prime}} \rceil + 6 + \log\epsilon^{- 1} +  (\lceil \log{M^{\prime}} \rceil - \lceil \log{M^{\prime \prime}} \rceil)$\\
Final system space & & $2\lceil\log M^{\prime}\rceil+ \log\epsilon^{- 1}$ & &   &  &$2\lceil\log M^{\prime\prime}\rceil+ \log\epsilon^{- 1}$  &\\
\hline
\end{tabular}
\end{table*}
\par Tab. \ref{tab:table2} presents the memory complexity of our method compared with two related works \cite{kerenidis2019QCNN, li2020QCNN} and its classical counterpart. The storage requirements of the model in Ref. \cite{kerenidis2019QCNN} amount to $ {M^{\prime}}^{2}N^{2} (4\lceil \log ({M^{\prime}}N) \rceil^{2}+ L) + N^{2} (4\lceil \log N \rceil^{2} + L) $, as it needs to expand each region traversed by the filter into a vector for storage and store the output of each layer in the QRAM \cite{QRAM, Giovannetti2008}. The working memory is $2\lceil \log(M^{\prime}N^{2})\rceil + 2 + \log\epsilon^{- 1}$, which is used for data loading and feature estimation. For Ref. \cite{li2020QCNN} which does not consider the pooling layer, two convolution layers are discussed here. It is similar to ours, requiring only the storage of information related to the original data scale, with a storage space of $M^{2} (2\lceil \log M \rceil^{2}+L) + N^{2} (2\lceil \log N \rceil^{2} + L) $ qubits. During working, additional qubits are needed to load data with a precision of $2^{-L}$ and to perform quantum arithmetic for the multiplication \cite{ruiz2017} of image information with kernel weight elements. In total, the working memory of Ref. \cite{li2020QCNN} is $2 \lceil \log M \rceil +2\lceil \log{M^{\prime}} \rceil + 6L +4+2\log\epsilon^{- 1}$. Furthermore, the memory requirement of the classical counterpart are $(M^{2} + N^{2} )L + ( {M^{\prime}}^{2} +{N^{\prime}}^{2})L $ and $M^{2} N^{2} + {M^{\prime}}^{2} {N^{\prime}}^{2}$, respectively. Obviously, the above quantum schemes can achieve a logarithmic reduction in workspace requirements compared with the classical counterpart, although requiring the same order of magnitude in storage memory. Hence, the proposed method demonstrates the exponential advantage of loading resources in quantum superposition.
\par Generally, the variables related to data size have the fact that $N < M^{\prime}\leq M$. In this case, the proposed method consumes fewer qubits than the other two quantum schemes when the precision of the information $L = \log\epsilon^{- 1} = \lceil\log M\rceil$. It implies that the proposed method can be implemented with a reduced quantum space compared with the approach in Ref. \cite{li2020QCNN}, while allowing for more flexible selection of stride and sliding window size. Moreover, our method mitigates the impact of sliding window size (i.e., the product of $N^{2}$) on the storage of data information, which is a limitation found in Ref. \cite{kerenidis2019QCNN}. These results indicate that our method may be more easily implemented on emerging quantum computing devices.
\begin{table}
\centering
\caption{\label{tab:table2}The memory resources comparison between the proposal and others.}
\renewcommand\arraystretch{1.5}
\begin{tabular}{m{0.08\textwidth}<{\centering} m{0.20\textwidth}<{\centering} m{0.17\textwidth}<{\centering}}
\hline
Method &Storage Memory &Working Memory \\
\hline
Kerenidis et. al \cite{kerenidis2019QCNN} &${M^{\prime}}^{2}N^{2} (4\lceil \log ({M^{\prime}}N) \rceil^{2}+ L) + N^{2} (4\lceil \log N \rceil^{2} + L) $ & $2\lceil \log(M^{\prime}N^{2})\rceil + 2 + 2\log\epsilon^{- 1}$ \\
Li et. al \cite{li2020QCNN} &$M^{2} (2\lceil \log M \rceil^{2}+L) + N^{2} (2\lceil \log N \rceil^{2} + L) $ & $2 \lceil \log M \rceil +2\lceil \log(M^{\prime}) \rceil + 6L +4+2\log\epsilon^{- 1}$\\
This paper & $M^{2} (2\lceil \log M \rceil^{2}+L) + N^{2} (2\lceil \log N \rceil^{2} + L) $ & $4\lceil \log(M) \rceil+4\lceil \log(M^{\prime}) \rceil   + 6 +2\log\epsilon^{- 1}$ \\
Classical counterpart &$(M^{2} + N^{2} )L + ( {M^{\prime}}^{2} +{N^{\prime}}^{2})L $ &$M^{2} N^{2} + {M^{\prime}}^{2} {N^{\prime}}^{2}$ \\
\hline
\end{tabular}
\end{table}
\section{\label{sec:5}Experiments}
\par In this section, we perform numerical simulations of QCNNFS using the Qiskit simulation platform. It offers some frequently used quantum gates and provides small-scale quantum system simulation services, making it an ideal tool for simulating a limited number of quantum systems and transformations \cite{aleksandrowicz2019qiskit}. The experimental setting is as follows.
\par In the experiments, we employed the handwritten digit images from the MNIST dataset. Due to the limitations of the Qiskit simulator and device, the bilinear interpolation algorithm \cite{keys1981cubic} is performed to scale the images. This reduces every image from $28 \times 28$ pixels to $4 \times 4$. Subsequently, each image is flattened into a one-dimensional vector of length $16$ and normalized according to Sec. \ref{sec:3.0}. To evaluate the performance of the algorithm, we simulate image classification scenarios using the processed dataset. The experiments included two tests, with one focus on classifying handwritten digits ``$6$" and ``$9$", while the other on ``$3$" and ``$6$". These images are taken from $128$ samples in the MNIST test dataset.
\par Then, the Qiskit framework is used to construct a QCNNFS model. Given the resource constraints inherent in the Qiskit platform and devices, the circuits for the convolutional, pooling, and fully connected layers are implemented separately. These layers are connected by utilizing measurement and recoding to achieve communication among them. It can mitigate resources by efficiently reclaiming quantum resources after each layer, although this comes at the cost of increased runtime. The simulation is also performed with different stride settings, and the results are shown in Tab. \ref{tab:table3}. The results demonstrate the effectiveness of our method for classification on the MNIST dataset. The classification accuracy for digits ``$6$" and ``$9$" reached an optimal $96.88\%$ when the strides were set to $(2,1)$. For digits ``$3$" and ``$6$", the optimal classification accuracy of $92.97\%$ is found when the strides were set to $(1,1)$. These results indicate that different datasets exhibit sensitivity to stride configurations and that stride adjustments play a critical role in the algorithm. Consequently, the flexibility of the proposed algorithm in stride settings is meaningful.
\begin{table}
\centering
\caption{\label{tab:table3}The classification accuracy of the proposed algorithm at different stride settings. Here, $(s , s^{\prime} )$ is represented as the stride of the convolution layer is $s$, while the stride of the pooling layer is $s^{\prime}$. The filter size for the convolutional and pooling layers is set to $2 \times 2$ in each experiment.}
\renewcommand\arraystretch{1.5}
\begin{tabular}{m{0.15\textwidth}<{\centering} m{0.1\textwidth}<{\centering} m{0.1\textwidth}<{\centering} m{0.1\textwidth}<{\centering}}
\hline
\multirow{2}{*}{Class label}  &\multicolumn{3}{c}{Stride $(s,s^{\prime})$} \\
\cline{2-4}
&$(1,1)$ &$(1,2)$ &$(2,1)$\\
\hline
``$6$" or ``$9$" &$89.06\%$ &$88.28\%$ &$96.88\%$\\
``$3$" or ``$6$" &$92.97\%$ &$89.84\%$ &$89.84\%$\\
\hline
\end{tabular}
\end{table}
\par
\begin{figure}
        \centering
        \subfloat[]{
            \includegraphics[height=0.3\textwidth,width=0.38\textwidth]{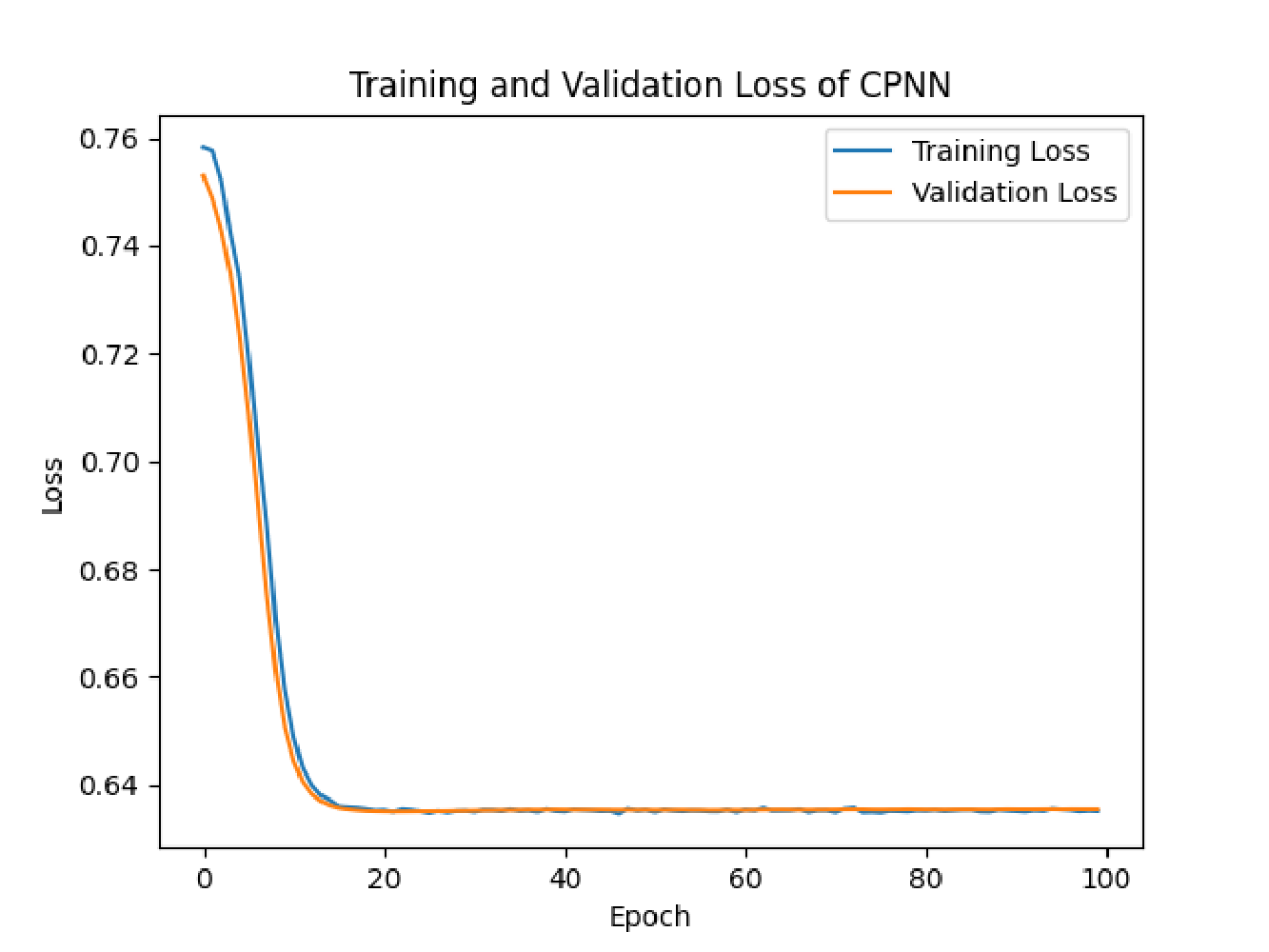}
        \label{fig:lossa}
        }
        \hfil
        \subfloat[]{
            \includegraphics[height=0.3\textwidth,width=0.38\textwidth]{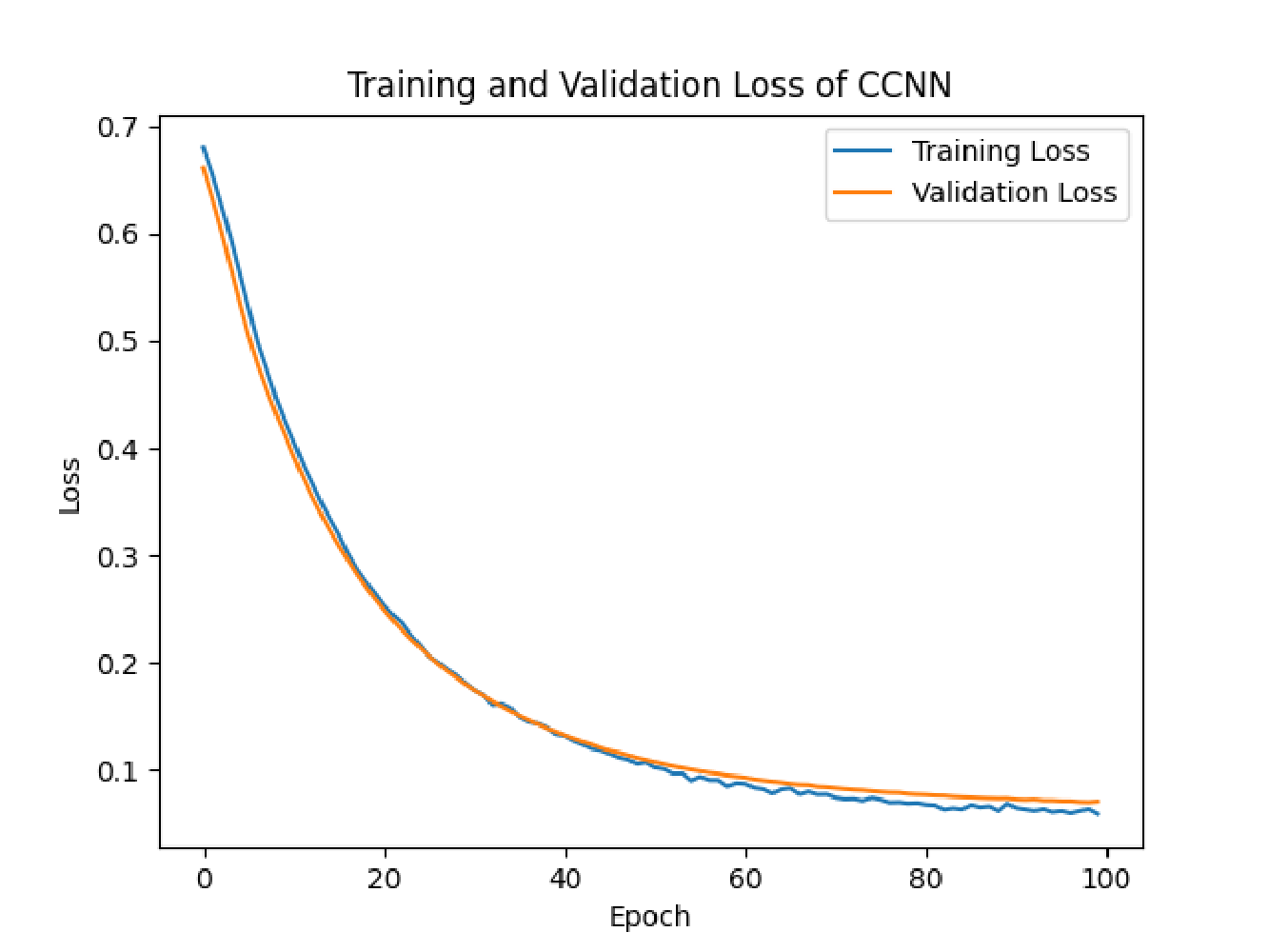}
        \label{fig:lossb}
        }
        \caption{Loss curves of two CNN architectures in binary classification task on MNIST dataset. (a) Training and validation losses of CPNN architecture. (b) Training and validation losses of CCNN architecture.}
        \label{fig:loss}
    \end{figure}

 Moreover, to explore the impact of different CNN architectures on performance, we conduct experiments using classical models with two distinct architectures. One uses a conventional architecture with a convolution layer, a pooling layer, and a fully connected layer (labeled as CPNN). This structure is also what the proposed scheme considers. Another follows Ref. \cite{li2020QCNN}, which omits the pooling layer and replaces it with an additional convolutional layer (named CCNN). Fig. \ref{fig:loss} illustrates the training and validation loss curves of them. In Fig. \ref{fig:lossa}, the CPNN architecture converges rapidly in fewer iterations, and its loss curve is smooth, with training and validation losses almost overlapping. In contrast, although the CCNN architecture achieves lower loss values (Fig. \ref{fig:lossb}), it shows fluctuations as epochs increases, particularly between epochs $70$ and $100$. The validation loss gradually exceeds the training loss, suggesting a risk of overfitting. Therefore, the CCNN architecture considered in this paper has stronger potential application trends in terms of stability and generalization ability.
\section{\label{sec:6}Conclusions}
\par In this paper, a novel algorithm for implementing convolutional neural networks using quantum primitive gates is presented. It aims to offer an efficient solution to the challenges of time efficiency and memory overhead faced by classical computing when handling large-scale data. Compared with the classical counterpart, QCNNFS requires only a logarithmic workspace and achieves exponential acceleration at the data scale. Additionally, the proposed method extends the potential applications of quantum CNN, particularly for tasks that require flexible stride adjustment. The analysis shows that the proposed method can achieve quantum acceleration with fewer quantum resources, compared with the related works. QCNNFS also avoids the extra memory overhead of increasing the sliding window size. By simulating handwritten digital images of the MNIST dataset on the Qiskit platform, we verified the effectiveness of QCNNFS. Finally, the influence of the choice of two CNN architectures on the testing is discussed, which shows that the architecture of the proposed algorithm has stronger stability in classification performance.
\par Certainly, some intriguing aspects for future research warrant further exploration. For instance, while the proposed algorithm has already optimized the consumption of qubits, investigating how to enhance resource efficiency further remains a valuable area for detailed investigation.

\section*{Acknowledgments}
This work was supported by National Natural Science Foundation of China (Grants No. 62171131, 61976053, and 61772134), Fujian Province Natural Science Foundation (Grant No. 2022J01186 and 2023J01533), and Innovation Program for Quantum Science and Technology (Grant No. 2021ZD0302901).

\bibliography{ref}

\end{document}